\documentstyle[preprint,aps]{revtex}
\begin{document}
\preprint{\begin{tabular}{l}
\hbox to \hsize{1999 March \hfill BROWN-HET-1141}\\[-3mm]
\hbox to \hsize{ \hfill KIAS-P99004}\\[5mm] \end{tabular} }
\bigskip

\title
{Calculability of Quark Mixing Parameters from General Nearest Neighbor Interaction Texture Quark Mass Matrices}
\author{David Dooling and Kyungsik Kang}

\address{Department of Physics\\
Brown University, Providence RI 02912, USA}

\maketitle

\begin{abstract}
\tightenlines
We perform an analysis of general quark mass matrices in the general nearest neighbor interaction form.
Excellent agreement with experiment is realized with this general texture, which is neither hermitian nor real-symmetric.
We then propose a new class of quark mass matrices that contain no additional parameters other than the quark masses themselves, and thus possess calculability, i.e., ensure a relationship between the six quark masses and four flavor-mixing parameters of the Standard Model. 
\end{abstract}
\tightenlines
\narrowtext

\section{\rm\bf  Introduction}

A fundamental explanation of the flavor-mixing matrix, the fermion masses and their hierarchical structure persists to be one of the most challenging and outstanding problems of particle physics today.
Within the standard model (SM) the fermion masses, the three flavor-mixing angles and the CP violating phase are free parameters and no relation exists among them.
However, the expectation that ``low-energy'' quantities which can be computed in the SM should remain finite as the masses of intermediate particles go to infinity leads us to suspect that such a relation does hold.
For example, $\Delta M (K_{s}^{o}-K_{L}^{o})$ diverges as $m_{t} \rightarrow \infty$.
Since the contribution of the intermediate quark is always multiplied by a flavor-mixing matrix element, a relation between quark masses and flavor-mixing parameters could conspire to guarantee that the contribution to the low-energy quantity always remains finite.

As an attempt to derive a relationship between the quark masses and flavor-mixing hierarchies, mass matrix ans\"{a}tze based on flavor democracy with a suitable breaking so as to allow mixing between the quarks of nearest kinship via nearest neighbor interactions (NNI) was suggested about two decades ago \cite{1}.
These early attempts are the first examples of ``strict calculability''; i.e., mass matrices such that all flavor-mixing parameters depend solely on, and are determined by, the quark masses.
But the simple symmetric NNI texture leads to the experimentally violated inequality $M_{top} < 110$ GeV, prompting consideration of a less restricted form for the mass matrices so as to still achieve calculability, yet be consistent with experiment \cite{2}.

It was later shown that the texture structure of these early ans\"{a}tze for quark mass matrices, the texture structure of the NNI mass matrices, holds in general.
Branco et al. have demonstrated that if one does not impose the assumption of hermiticity, then the NNI texture structure contains no physical assumptions \cite{3}.
For three fermion generations, one may consider without loss of generality quark mass matrices of the NNI form.
This texture structure serves as the general starting point from which additional constraints on the mass matrix elements can be imposed in order to achieve calculability.


In this paper we analyze general quark mass matrices in a modified NNI form to be described in the next section.
This form is free from physical content, as it corresponds to a particular choice of weak basis in the right-handed chiral quark sector.
We perform a numerical fit to the most recent measurement of the Cabibbo-Kobayashi-Maskawa (CKM) matrix, $V_{CKM}$.
The results are in excellent agreement with the observables.
Finally, we propose a new mass matrix ans\"{a}tze based on our numerical results, to ensure some relation among the six quark mass masses and the four observable flavor-mixing parameters, thus reducing the number of free parameters in the SM.

Our paper is organized as follows.
In Section II we review the NNI form of mass matrices, pointing out a subtlety that has been overlooked in previous works.
In Section III, we present the quark mass matrices to be analyzed as well as the resulting flavor-mixing matrix.
We present the observables used in our fitting procedure as well as the resultant unitary triangle (UT) parameters.
Guided by the numerics of the general case, we postulate quark mass matrices that ensure calculability.

\section{General NNI mass matrices}
In a given generic gauge theory,  flavor eigenstates do not necessarily coincide with the mass eigenstates which can be written in terms of combinations of flavor eigenstates.
The mass matrix in a given electric charge charge sector is not necessarily diagonal in flavor space and involves the coupling between the left-handed and right-handed chiral states of different flavor. 
It can be shown that in the SM with three generations of fermions, one may perform a biunitary transformation of the quark mass matrices that leaves both the quark mass spectrum and the flavor-mixing parameters unchanged, where the new mass matrices in both the up-type and down-type quark sectors are of the NNI form and are in general neither hermitian nor symmetric \cite{3}.
We stress that this form is merely the consequence of choosing a particularly convenient weak basis that allows us to eliminate some of the redundant free parameters of the SM and has no physical consequences.

Starting from initial completely general complex matrices $M_{(u,d)}^{\prime}$, a necessary condition for this biunitary transformation to yield the NNI form is the existence of a solution to the eigenvalue equation
\begin{equation}
\left( M_{u}^{\prime}M_{u}^{\prime \dagger}+kM_{d}^{\prime}M_{d}^{\prime\dagger} \right)_{ji}U_{i2} = \lambda U_{j2}
\end{equation}
where $k$ is initially assumed to be an arbitrary complex constant.
The NNI mass matrices are then given by

\begin{equation}
M_{u} = \left( \begin{array}{ccc} 
0 & A_{12}e^{i\theta_{1}} & 0 \\
A_{21}e^{i\theta_{2}} & 0 & A_{23}e^{i\theta_{3}} \\
0 & A_{32}e^{i\theta_{4}} & A_{33}e^{i\theta_{5}}
\end{array} \right)
\end{equation}
\begin{equation}
M_{d} = \left( \begin{array}{ccc}
0 & B_{12}e^{i\theta_{1}^{\prime}} & 0 \\
B_{21}e^{i\theta_{2}^{\prime}} & 0 & B_{23}e^{i\theta_{3}^{\prime}} \\
0 & B_{32}e^{i\theta_{4}^{\prime}} & B_{33}e^{i\theta_{5}^{\prime}}
\end{array} \right)
\end{equation}
\begin{equation}
M_{(u,d)} = U^{\dagger}M_{(u,d)}^{\prime}V_{(u,d)}
\end{equation}
where explicit forms for $U$ and $V_{(u,d)}$, corresponding to the left-handed and right-handed chiral quark rotations, respectively, are given in \cite{3}.
Note that $U$, the left-handed chiral quark rotation matrix, is common to both the up and down quark sectors.
It is the freedom of performing an unitary transformation on the triplet of right-handed quark fields in both the up and down quark sectors, i.e. the freedom associated with the $V_{(u,d)}$, without affecting the quark masses or the flavor-mixing parameters, that enables this texture structure to be possible while still remaining completely general.
We can rewrite Eq. (1) for arbitrary complex $k$ as
\begin{displaymath}
A_{21}^{2}+A_{23}^{2}+k(B_{21}^{2}+B_{23}^{2}) = \lambda
\end{displaymath}
\begin{displaymath}
k_{real}\left[\cos(\beta)\sin(\alpha)-\sin(\beta)\cos(\alpha)\right]-k_{imag}\left[\sin(\beta)\sin(\alpha)+\cos(\beta)\cos(\alpha) \right] = 0
\end{displaymath}
where $\beta \equiv \theta_{5}^{\prime} - \theta_{3}^{\prime}$ and $\alpha \equiv \theta_{5} - \theta_{3}$.
The first equation exhibits the functional dependence of $\lambda$ on $k$ and provides no restriction on $k$.
It may appear that when choosing a weak-basis that gives NNI forms for the quark mass matrices, one has two arbitrary degrees of freedom corresponding to $k_{real}$ and $k_{imag}$.
But in fact this two-fold degree of freedom does not exist; $k$ must be either purely real or purely imaginary.

 For both $k_{real}$ and $k_{imag}$ nonzero, we arrive at the inconsistency $\tan(\beta) = -\cot(\beta)$.
Inspection of Eq. (1) shows that if $k$ is purely real, we are guaranteed a solution to (1) because the operator on the left hand side is Hermitian and has three linearly independent eigenvectors.
 However, this guarantee does not hold if $k$ is purely imaginary.
Therefore, we arrive at the condition $k$ is purely real, which results in $\alpha = \beta$.
Thus the number of parameters in our flavor-mixing matrix is reduced to eleven from twelve. 

The above mass matrices with their texture structure and the relationship $\alpha = \beta$ are the most general ones that exhibit the NNI texture and have the fewest number of independent parameters.
At this stage no physical inputs have been introduced, and the resulting flavor-mixing matrix has five more parameters than are necessary to acheive strict calculability.

\section{Flavor-mixing matrix and numerical fit to observables}

We use the rephasing freedom of the quark-fields to minimize the number of parameters entering into the quark flavor-mixing matrix.
We can re-write Eq. (2)
\begin{equation}
M_{u} = \left( \begin{array}{ccc}
e^{i(\theta_{1}-\theta_{4})} & 0 & 0 \\
0 & e^{i(\theta_{3}-\theta_{5})} & 0 \\
0 & 0 & 1  \end{array} \right) \left( \begin{array}{ccc}
0 & A_{12} & 0 \\
A_{21} & 0 & A_{23} \\
0 & A_{32} & A_{33} \end{array} \right) \left(\begin{array}{ccc}
e^{i(\theta_{5}-\theta_{3}+\theta_{2})} & 0 & 0 \\
0 & e^{i(\theta_{4})} & 0 \\
0 & 0 & e^{i(\theta_{5})} \end{array} \right) \equiv P_{(u)L}\tilde{M_{u}}P_{(u)R}
\end{equation}
with an analogous expression for $M_{d}$, and where $\theta_{3}^{\prime} - \theta_{5}^{\prime} = \theta_{3} - \theta_{5}$.
The flavor-mixing matrix is written in terms of the unitary matrices $U_{(u)L}$ and $U_{(d)L}$ that diagonalize $M_{u}M_{u}^{\dagger}$ and $M_{d}M_{d}^{\dagger}$:
\begin{displaymath}
U_{(u)L}M_{u}M_{u}^{\dagger}U_{(u)L}^{\dagger} \equiv \mbox{diag}(m_{u}^{2},m_{c}^{2},m_{t}^{2})
\end{displaymath}
\begin{displaymath}
U_{(d)L}M_{d}M_{d}^{\dagger}U_{(d)L}^{\dagger} \equiv \mbox{diag}(m_{d}^{2},m_{s}^{2},m_{b}^{2})
\end{displaymath}
From the above expressions for $M_{u,d}$, we see that 
\begin{equation}
M_{u}M_{u}^{\dagger} = P_{L}\tilde{M_{u}}\tilde{M_{u}}^{T}P_{L}^{\dagger}
\end{equation}
Because $\tilde{M_{u}}\tilde{M_{u}}^{T}$ is a real symmetric matrix, it can be diagonalized by a real orthogonal matrix $R_{u}$:
\begin{equation}
R_{u}\tilde{M_{u}}\tilde{M_{u}^{T}}R_{u}^{T} = \mbox{diag}(m_{u}^{2},m_{c}^{2},m_{t}^{2}).
\end{equation}
The flavor-mixing matrix can then be expressed as 
\begin{equation}
V_{f-m} = U_{(u)L}U_{(d)L}^{\dagger} = R_{u}P_{(u)L}^{\dagger}P_{(d)L}R_{d}^{T}
\end{equation}
\begin{equation}
V_{f-m} = \left( \begin{array}{ccc}
\leftarrow & \vec{v_{u}} & \rightarrow \\
\leftarrow & \vec{v_{c}} & \rightarrow \\
\leftarrow & \vec{v_{t}} & \rightarrow
\end{array} \right) \left( \begin{array}{ccc}
e^{i\theta} & 0 & 0 \\
0 & 1 &  0 \\
0 & 0 & 1 
\end{array} \right) \left( \begin{array}{ccc}
\uparrow & \uparrow & \uparrow \\
\vec{v_{d}} & \vec{v_{s}} & \vec{v_{b}} \\
\downarrow & \downarrow & \downarrow
\end{array} \right)
\end{equation}
where the $\vec{v}_{u,c,t}$ are the normalized eigenvectors of $\tilde{M_{u}}\tilde{M_{u}}^{\dagger}$ with eigenvalues $m_{u}^{2}, m_{c}^{2}$ and $m_{t}^{2}$ and similarly for the $\vec{v}_{d,s,b}$ \cite{4,5,6}.
Note that only this single phase $\theta \equiv \theta_{1}-\theta_{4}-\theta_{1}^{\prime}+\theta_{4}^{\prime}$ enters the flavor-mixing matrix and is responsible for CP violation.
Thus, it is sufficient for only one of the four non-vanishing elements coming from the second columns of $M_{u}$ and $M_{d}$ to be complex in order to provide a general parametrization of flavor-mixing and CP violation.
For example, one may choose $\theta_{1}$ to be non-zero while all other phases vanish.
At this stage, the mass matrices contain eleven independent parameters.
 
We must specify the energy scale at which we are evaluating the mass matrices.
We use the light quark masses \cite{7,8} $m_{u} = 4.9 \pm 0.53$ MeV, $m_{d} = 9.76 \pm 0.63$ MeV, and $m_{s} = 187 \pm 16$ MeV, and the heavy quark masses $m_{c} = 1.467 \pm 0.028$ GeV, $m_{b} = 6.356 \pm 0.08$ GeV and $m_{t} = 339 \pm 24$ GeV all of which correspond to the masses at a modified minimal subtraction $(\overline{MS})$ renormalization point of 1 GeV.
Choosing the energy scale to be 1 GeV we use the invariance of the trace, minor determinants and determinant of $\tilde{M}_{u(d)}\tilde{M}_{u(d)}^{T}$ under a similarity transformation to relate the $\tilde{M}_{u(d)ij}$ to the $m_{i}^{2}$ in each quark sector. 
These six relations comprise the first six terms of our $\chi_{tot}^{2}$.


In addition, we fit to the following $V_{CKM}$ observables \cite{9}: $|V_{ud}| = 0.9740 \pm 0.001$, $|V_{us}| = 0.2196 \pm 0.0023$, $|V_{cd}| = 0.224 \pm 0.016$, $|V_{cs}| = 1.04 \pm 0.16$, $|V_{cb}| = 0.0395 \pm 0.0017$, and $\frac{|V_{ub}|}{|V_{cb}|} = 0.08 \pm 0.02$.

Our method is to fit to the experimental data and compute the corresponding $\chi_{tot}^{2} \equiv \sum_{i=1}^{12} \chi_{i}^{2}$ with one degree of freedom arising from twelve experimental values minus eleven independent input parameters.

The following parameters yield a $\chi_{tot}^{2}$ of only 0.180:

\vspace{.5cm}

Table 1
\vspace{.5cm}

\begin{tabular}[t]{|l|l||l|l||} \hline 
 Parameter & Value [MeV] & Parameter & Value [Mev]  \\
\hline \hline
$A_{12}$ & 976.99 $\pm$ 0.9818 & $B_{12}$ & $44.149 \pm 0.1959$ \\
\hline
$A_{21}$ & 101.07 $\pm$ 0.975 & $B_{21}$ & $50.532 \pm 1.406$ \\
\hline
$A_{23}$ & -1465.5 $\pm$ 0.5671 & $B_{23}$ & $304.19 \pm 3.2851$ \\
\hline
$A_{32}$ & -3.3811 $\times 10^{5}$ $\pm$ 0.9675 & $B_{32}$ & $-3644.1 \pm 0.9721$ \\
\hline
$A_{33}$ & -24678 $\pm$ 0.9679 & $B_{33}$ & $-5203.4 \pm 0.9693$ \\  
\hline
\end{tabular}
\vspace{.5cm}

The single phase $\theta$ that minimizes $\chi_{tot}^{2}$ is -0.716 $\pm$ 0.0989 radians.
The above parameters also make predictions concerning the mixing of the top quark as follows: $\frac{|Vtb|^{2}}{|V_{td}|^{2}+|V_{ts}|^{2}+|V_{tb}|^{2}} = 0.9984$, $|V_{tb}^{\ast}V_{td}| = 0.009422$ and $\frac{|V_{td}|}{|V_{ts}|} = 0.2479$.
The first two predictions above are in excellent agreement with the latest Particle Data Group values \cite{9} , while our value of $\frac{|V_{td}|}{|V_{ts}|}$ is slightly higher than that predicted in \cite{10,11}.

The numerically determined mass matrices $\tilde{M_{u}}, \tilde{M_{d}}$ from above display some interesting properties worth noting at this point.
One can immediately see that they are far from being symmetric, as $|A_{12}|$ differs markedly in magnitude from $|A_{21}|$, etc.
Of particular interest is the difference between the up and down quark sectors vis-a-vis the 3-3 element.
Historically, mass matrix ans\"{a}tze previously proposed in the spirit of calculability have written the 3-3 element as a small pertubation about $|m_{t,b}|$; i.e. as $|m_{3} - \epsilon|$, where epsilon is some small parameter.
Inspection of the above parameters shows that this description works fairly well for the down sector, but is not at all applicable to the up quark sector.
In the up quark sector, it is the magnitude of the 3-2 element that differs only slightly from $m_{3} = m_{t}$. 
This numerical result and observed structure is a consequence of the top quark being so heavy.

Using just the central values of the quark masses at the 1 GeV scale, the absolute values of the elements of the flavor-mixing matrix elements are:

\begin{displaymath}
|V_{CKM}| = \left( \begin{array}{ccc}
0.974007 & 0.219610 & .00316079 \\
0.223856 & 0.975401 & .0395023 \\
.00942966 & .0380247 & 0.999214 \end{array} \right)
\end{displaymath}

To impose unitarity we need to go through one additional step after varying our input parameters and fitting to the above experimental values.
In our flavor-mixing matrix, we have in the matrices
\begin{displaymath}
\left( \begin{array}{ccc}
\longleftarrow & \vec{v_{u}} & \longrightarrow \\
\longleftarrow & \vec{v_{c}} & \longrightarrow \\
\longleftarrow & \vec{v_{t}} & \longrightarrow \end{array} \right)~~~\mbox{and}~~~ \left( \begin{array}{ccc}
\uparrow & \uparrow & \uparrow \\
\vec{v_{d}} & \vec{v_{s}} & \vec{v_{b}} \\
\downarrow & \downarrow & \downarrow \end{array} \right)
\end{displaymath}
explicit expressions involving the $m_{i}^{2}$, the six central values of the quark masses squared at the 1 GeV scale.
The real symmetric matrices $\tilde{M_{u}}\tilde{M_{u}}^{T}$ and $\tilde{M_{d}}\tilde{M_{d}}^{T}$ will not have these central values as their exact eigenvalues unless the fit is perfect; i.e. unless $\sum_{i=1}^{6} \chi_{i}^{2} = 0$.

To ensure unitarity, after having performed the fit to the experimental observations, we evaluate the eigenvalues of $\tilde{M_{u}}\tilde{M_{u}}^{\dagger}$ and $\tilde{M_{d}}\tilde{M_{d}}^{\dagger}$ to find the correct $m_{i}^{2}$ we should use in the $\vec{v_{i}}$ and then re-evaluate our expression for $V_{f-m}$, thus ensuring an unitary flavor-mixing matrix.

Although unitary, our derived flavor-mixing matrix contains unphysical phases that must be removed in order to perform an unitary triangle analysis.
After finding the mass matrices parameters that yield the minimum $\chi^{2}$, we put the flavor-mixing matrix into the standard CKM representation advocated by the Particle Data Group and then into an improved Wolfenstein parametrization from which we find the unitary triangle.
%




The improved Wolfenstein parametrization can be obtained from the standard CKM representation with the following identifications \cite{9,12}:
\begin{displaymath}
\lambda \equiv s_{12} = \sin{\theta_{12}},  A\lambda^{2} \equiv s_{23} = \sin{\theta_{23}} , A\lambda^{3}(\rho - i\eta) \equiv s_{13}e^{-i\delta} = \sin{\theta_{13}}e^{-i\delta}
\end{displaymath}
with modified Wolfenstein parameters
\begin{displaymath}
\overline{\rho} \equiv \rho(1-\frac{\lambda^{2}}{2}), \overline{\eta} \equiv \eta(1-\frac{\lambda^{2}}{2})
\end{displaymath} 

With the above parameter values and using the true numerical eigenvalues of the resulting $\tilde{M}_{(u,d)}\tilde{M}_{(u,d)}^{T}$, we arrive at the following values for the flavor-mixing parameters:

\vspace{0.5cm}

Table 2
\vspace{.5cm}

\begin{tabular}{|l|l||l|l||} \hline
Parameter & Value [rad] & Parameter & Value \\
\hline \hline
$\theta_{13}$ & $3.160065 \times 10^{-3}$ &  $\overline{\eta}$ & $0.3444959$  \\
\hline
$\theta_{12}$ & $0.22825323$ & $\overline{\rho}$ & $0.00422775$ \\
\hline
$\theta_{23}$ & $3.95070522 \times 10^{-2}$ &  $\alpha$ & $71.61961^{\circ}$ \\
\hline
$\delta$ & $1.5585246$ & $\beta$ &  $19.083502^{\circ}$ \\
\hline
$\lambda$ & $.22627623$ & $\gamma$ & $89.296884^{\circ}$ \\
\hline
A  & $.771433972$ & $J$ & $2.748573468 \times 10^{-5}$ \\
\hline
\end{tabular}

\vspace{0.5cm}
$J$ is the Jarlskog invariant \cite{13,5}, a rephasing invariant of the mixing matrix, and is given in the standard representation of the quark flavor-mixing matrix as $J = s_{12}s_{13}s_{23}c_{12}c_{13}^{2}c_{23}s_{\delta} = 2.748573468 \times 10^{-5}$.

After imposing unitarity as described above, the absolute values of $V_{CKM}$ become:

\begin{displaymath}
|V_{CKM}| = \left( \begin{array}{ccc}
0.974058 & 0.226275 & .00316006 \\
0.226101 & 0.973303 & .0394979 \\
.00941607 & .0384891 & .999215 \end{array} \right)
\end{displaymath}

As alluded to in the introduction, one of the most appealing features of the NNI mass matrices is the large number of texture zeroes.
One may take advantage of this texture structure to eliminate three of the five real parameters in each sector, expressing them solely in terms of the $m_{i}^{2}$, up to a sign ambiguity associated with the square root branches.
That is, we may use the invariance of the characteristic equation under a similiarity transformation to express $A_{23}, A_{32},$ and $A_{33}$ in terms of $m_{u}^{2}, m_{c}^{2},$ and $m_{t}^{2}$, and similarly for the down quark sector.
Choosing $A_{12}$ and $A_{12}$ to be the as yet undetermined parameters in the up-quark mass matrix, and using the branch information revealed by Table 1, one easily verifies the following relations: $A_{33} = -\frac{m_{u}m_{c}m_{t}}{A_{12}A_{21}}$, $A_{32} = -\sqrt{\frac{b+\sqrt{b^{2}+4c}}{2}}$, $A_{23} = -\sqrt{m_{u}^{2}+m_{c}^{2}+m_{t}^{2}-A_{12}^{2}-A_{21}^{2}-A_{32}^{2}-\frac{m_{u}^{2}m_{c}^{2}m_{t}^{2}}{A_{12}^{2}A_{21}^{2}}}$,
where
$b \equiv m_{u}^{2}+m_{c}^{2}+m_{t}^{2}-2A_{12}^{2}-\frac{m_{u}^{2}m_{c}^{2}m_{t}^{2}}{A_{12}^{2}A_{21}^{2}}$ and $c \equiv A_{12}^{2}\left( m_{u}^{2}+m_{c}^{2}+m_{t}^{2}-A_{12}^{2} \right) + \frac{m_{u}^{2}m_{c}^{2}m_{t}^{2}}{A_{12}^{2}}-m_{u}^{2}m_{c}^{2}-m_{u}^{2}m_{t}^{2}-m_{c}^{2}m_{t}^{2}$.

Similarly, for the down-quark sector, choosing $B_{12}$ and $B_{21}$ to be the parameters as yet undetermined by the invariance relations, one finds the following equalities: $B_{33} = -\frac{m_{d}m_{s}m_{b}}{B_{12}B_{21}}$, $B_{32} = -\sqrt{\frac{d+\sqrt{d^{2}+4e}}{2}}$, $B_{23}=\sqrt{m_{d}^{2}+m_{s}^{2}+m_{b}^{2}-B_{12}^{2}-B_{21}^{2}-B_{32}^{2}-\frac{m_{d}^{2}m_{s}^{2}m_{b}^{2}}{B_{12}^{2}B_{21}^{2}}}$,
where $d \equiv m_{d}^{2}+m_{s}^{2}+m_{b}^{2}-2B_{12}^{2}-\frac{m_{d}^{2}m_{s}^{2}m_{b}^{2}}{B_{12}^{2}B_{21}^{2}}$ and $e \equiv B_{12}^{2}\left( m_{d}^{2}+m_{s}^{2}+m_{b}^{2}-B_{12}^{2} \right) + \frac{m_{d}^{2}m_{s}^{2}m_{b}^{2}}{B_{12}^{2}}-m_{d}^{2}m_{s}^{2}-m_{d}^{2}m_{b}^{2}-m_{s}^{2}m_{b}^{2}$.


Using the above mass matrices with their associated five total degrees of freedom corresponding to $A_{12},A_{21},B_{12},B_{21}$ and $\theta$, and the six experimental values of $|V_{CKM}|$ observables as before, we obtain Table 3, with a $\chi_{tot}^{2}$ of 2.182. 
This $\chi_{tot}^{2}$ is not nearly as good as in the previous case, but the simple explanation for this apparent decline in numerical confidence is just a reflection of the slight discrepancy between unitarity and the present experimental values. 
In the previous case, the flavor-mixing matrix could deviate from unitarity to satisfy the experimental constraints; i.e., $|V_{cs}|$ having a central value that in itself violates unitarity did not do violence to the minimization of $\chi_{tot}^{2}$ because the requirements of unitarity were not automatically satisfied.
In the current situation, unitarity \textit{is} automatically imposed, and so the increase in $\chi_{tot}^{2}$ is to be expected. 
 The predicted $|V_{CKM}|$ all lie within the ranges preferred by the Particle Data Group analysis, and as such, any set of parameters that fulfill this requirement of agreement with the ranges listed therein is deemed  ``acceptable'', regardless of $\chi_{tot}^{2}$ values.
\begin{displaymath}
|V_{CKM}| = \left( \begin{array}{ccc}
.975244 & .22111 & .00314624 \\
.220939 & .974488 & .0394887 \\
.00924305 & .0385204 & .999215 \end{array} \right)
\end{displaymath}
Implicit in this prejudicial ``acceptance'' is the belief that only three generations of quarks exist and that future experiments will bring the experimental values into closer agreement with the constraints imposed by unitarity.

\vspace{0.5cm}
\newpage

Table 3
\vspace{.5cm}

\begin{tabular}{|l|l||} \hline
Parameter & Value \\
\hline \hline
$A_{12}$ & $1229.3 \pm 32.38$ \\
\hline
$A_{21}$ & $139.82 \pm .091$ \\
\hline
$B_{12}$ & $45.872 \pm 3.4573$ \\
\hline
$B_{21}$ & $48.797 \pm .0433$ \\
\hline
$\theta$ & $-0.67806 \pm .009657$ \\
\hline
\end{tabular}

\vspace{0.5cm}


In addition, the standard representation parameters as well as the Wolfenstein parameters and Jarlskog invariant $J$ are predicted to be:

\vspace{0.5cm}

Table 4
\vspace{.5cm}

\begin{tabular}{|l|l||l|l||} \hline
Parameter & Value & Parameter & Value \\
\hline \hline
$\theta_{13}$ & $3.146249 \times 10^{-3}$ & $\overline{\eta}$ & $.3515246$ \\
\hline
$\theta_{12}$ & $0.2229535$ & $\overline{\rho}$ & $.001394$ \\
\hline
$\theta_{23}$ & $3.9499098 \times 10^{-2}$ & $\alpha$ & $70.83438^{\circ}$ \\
\hline
$\delta$ & $1.5668302$ & $\beta$ & $19.39285^{\circ}$ \\
\hline
$\lambda$ & $0.2211108$ & $\gamma$ & $89.77276^{\circ}$ \\
\hline
$A$ & $.80770906$ & $J$ & $2.67698466 \times 10^{-5}$ \\
\hline
\end{tabular} 
\vspace{0.5cm}

Because the mass matrices squared have the correct eigenvalues, the free-parameters $A_{12}, A_{21}, B_{12}, B_{21}$ and $\theta$ represent the general parametrisation at the fundamental mass matrices in general.
To realize calculability, it is not enough to simply postulate some relation among this set of parameters \textit{independent} of the quark masses, as such a relation would only serve to relate the flavor-mixing parameters among themselves, and not provide the sought after relation between quark masses and flavor-mixing parameters. 

\section{Making New Mass Matrix Ans\"{a}tze}

The above mass matrices are completely general; to achieve calculability we must find expressions for $A_{12}, A_{21}, B_{12}, B_{21}$ and $\theta$ in terms of the $m_{i}$.
Guided by the parameter values in Tables 1 and 3, we postulate the following mass matrices in terms of the $m_{i}$:





\begin{displaymath}
\tilde{M_{u}} = \left( \begin{array}{ccc}
0 & A_{12} & 0 \\
A_{21} & 0 & -\sqrt{m_{u}^{2}+m_{c}^{2}+m_{t}^{2}-A_{12}^{2}-A_{21}^{2}-A_{32}^{2} -\frac{m_{u}^{2}m_{c}^{2}m_{t}^{2}}{A_{12}^{2}A_{21}^{2}}} \\
0 & -\sqrt{\frac{b+\sqrt{b^{2}+4c}}{2}} & -\frac{m_{u}m_{c}m_{t}}{A_{12}A_{21}} \end{array} \right)
\end{displaymath}
where $A_{12} = \left( \sqrt{\frac{m_{u}}{2}}+\sqrt{\frac{m_{c}}{2}} \right)^{2}$, $A_{21} = \sqrt{m_{u}m_{c}}$ and

\begin{displaymath}
\tilde{M_{d}} = \left( \begin{array}{ccc}
0 & B_{12} & 0 \\
B_{21} & 0 & \sqrt{m_{d}^{2}+m_{s}^{2}+m_{b}^{2}-B_{12}^{2}-B_{21}^{2}-B_{32}^{2}-\frac{m_{d}^{2}m_{s}^{2}m_{b}^{2}}{B_{12}^{2}B_{21}^{2}}} \\
0 & -\sqrt{\frac{d+\sqrt{d^{2}+4e}}{2}} & -\frac{m_{b}\sqrt{m_{d}m_{s}}}{B_{21}} \end{array} \right)
\end{displaymath}
where $B_{12} = \sqrt{m_{d}m_{s}}$, $B_{21} = m_{d} + \sqrt{m_{d}m_{s}}$ and $b,c,d$ and $e$ are defined as before.



Lastly, the single phase $\theta$ entering the flavor-mixing matrix is postulated to be $-\frac{B_{32}}{B_{33}}$.
We know that for only two quark generations there is no CP violating phase in the mixing matrix, so it is natural to expect, within the framework of calculability, that $\theta$ will involve ratios of elements in the mass matrices that only involve the mixings of the third generation quarks.
With these postulates for $\tilde{M_{u}}$, $\tilde{M_{d}}$ and $\theta$, $V_{f-m}$ is found from Eqns. (6-9).
This constitutes a new mass matrix ansatz that provides a calculable model of flavor-mixing and is in excellent agreement with the latest experimental results. 

This new ansatz predicts the following absolute values for $|V_{CKM}|$:

\begin{displaymath}
|V_{CKM}| = \left( \begin{array}{ccc}
.974427 & .224682 & .00307663 \\
.224513 & .973672 & .0394506 \\
.00923256 & .0384783 & .999217 \end{array} \right),
\end{displaymath}
  $\frac{|V_{ub}|}{|V_{cb}|} = 0.0779868$ and $\frac{|V_{td}|}{|V_{ts}|} = 0.239942$.
The $\chi_{tot}^{2}$ found from the six experimental $V_{CKM}$ measurements is found to be 5.2489.
Because we have no free parameters, there are six degrees of freedom.
Such a value for $\chi_{tot}^{2}$ corresponds to $\sim 70 \%$ confidence level.

The standard representation parameters, Jarlskog invariant $J$, and the modified Wolfenstein parameters are predicted to be:

\vspace{0.5cm}

Table 5
\vspace{.5cm}

\begin{tabular}{|l|l||l|l||} \hline
Parameter & Value & Parameter & Value \\
\hline \hline
$\theta_{13}$ & $3.0766324 \times 10^{-3}$ & $A$ & $0.781473947$ \\
\hline
$\theta_{12}$ & 0.226617779 & $\overline{\eta}$ & $0.3380146$ \\
\hline
$\theta_{23}$ & $3.94622992 \times 10^{-2}$ & $\overline{\rho}$ & $0.01469470$ \\
\hline
$\delta$ & 1.52735011 & $\alpha$ & $73.554460004^{\circ}$ \\
\hline
$J$ & $2.652858 \times 10^{-5}$ & $\beta$ & $18.93482465^{\circ}$ \\
\hline
$\lambda$ & 0.2246832 & $\gamma$ & $87.51071528^{\circ}$ \\
\hline
\end{tabular}

\vspace{0.5cm}

The flavor-mixing observables predicted from this calculability ansatz in Table 5 are in excellent agreement with the general result of Table 2.

The above mass matrices with their calculability property as well as the low $\chi_{tot}^{2}$ are very compelling arguments in favor of calculability in the quark sector, but naturally one would like to uncover at least a glimpse of the more fundamental theory beyond the SM that is the source of this calculability.
When one considers hermitian mass matrices, considerations of family permutation symmetry and its breaking in successive stages are suggestive explanations of the source of calculability, whereas with general NNI texture mass matrices, even though calculability still holds, it is not readily apparent what familiy symmetry, if any, is responsible.
In conclusion, we have elucidated an important point in the construction of NNI weak-eigenstate quark basis that has eluded some previous authors.
We have then performed an analysis of the experimental data to determine the mass matrix parameters, and have obtained values for the flavor-mixing and UT parameters that are in excellent agreement with previous analysis \cite{12}.
Finally, we have presented a new class of calculable mass matrices that also can explain all the experimental data and predict nearly identical values for the flavor-mixing and UT parameters as in the previous case.

\section{ACKNOWLEDGEMENTS}

We thank the hospitality of the School of Physics, Korea Institute of Advanced Study, where much of this work was completed.
DD would like to thank the NSF/KOSEF 1998 Summer Institute in Korea Program for finanacial support.
Support for this work was provided in part by U.S. Dept. of Energy Contract DE-FG-02-91ER40688-Task A.

\end{document}